\documentclass[%
 reprint,
 superscriptaddress,
 amsmath,amssymb,
 aps,
prl,
]{revtex4-2}

\usepackage{graphicx}
\usepackage{dcolumn}
\usepackage{bm}
\usepackage{hyperref}
\usepackage{url}
\hypersetup{hidelinks}
\usepackage[dvipsnames]{xcolor}
\usepackage{makecell}
\usepackage{physics}
\usepackage{upgreek}
\usepackage{float}
\usepackage{amsmath}
\usepackage{siunitx}
\usepackage{bbding}
\usepackage{multirow}
\usepackage{mathrsfs}

\begin{document}

\title{The nature of polar distortions in ferroelectrics}

\author{Hong Jian Zhao}
 \affiliation{Key Laboratory of Material Simulation Methods and Software of Ministry of Education, College of Physics, Jilin University, Changchun 130012, China}
 \affiliation{Key Laboratory of Physics and Technology for Advanced Batteries (Ministry of Education), College of Physics, Jilin University, Changchun 130012, China}
 \affiliation{International Center of Future Science, Jilin University, Changchun 130012, China}
\author{Laurent Bellaiche}
\affiliation{Smart Functional Materials Center, Physics Department and Institute for Nanoscience and Engineering, University of Arkansas, Fayetteville, Arkansas 72701, USA}
\author{Yanming Ma}
\affiliation{School of Physics, Zhejiang University, Hangzhou 310058, China}
 \affiliation{Key Laboratory of Material Simulation Methods and Software of Ministry of Education, College of Physics, Jilin University, Changchun 130012, China}

\begin{abstract}
Polar distortion, the collective off-center displacements of atoms, is a fingerprint of a ferroelectric that governs its properties and functionalities. Since the 1970s, the concepts of proper, improper and triggered ferroelectrics have been established to shed light on a diversity of polar distortion mechanisms. Such concepts assign a single nature to polar distortion and are helpful to interpret how polar distortions occur in conventional ferroelectrics such as barium titanate. However, applying these concepts to complex ferroelectrics (e.g., polar orthorhombic hafnia) is notoriously challenging and can yield highly controversial arguments. Here we resolve this issue by developing a tailor-made graph theory for clarifying the nature of polar distortions in complex ferroelectrics, which emphasizes that polar distortions in such ferroelectrics usually exhibit multiple natures among proper, improper and triggered characteristics. We demonstrate the robustness of our theory by working with perovsktie superlattices and polar orthorhombic hafnia (i.e., two representative cases). We successfully identify the mixed proper-improper nature in perovsktite superlattices and reconcile the controversy on polar orthorhombic hafnia by confirming its mixed trigger-improper nature. Our work will definitely lead to a revisitation of concepts in ferroelectric physics and provide opportunities for discovering novel ferroelectrics and related phenomena. 
\end{abstract}

\maketitle

\noindent
\textit{Introduction.} Ferroelectrics are advanced materials with broad technological applications (e.g., energy harvesters, sensors, actuators and non-volatile memories)~\cite{ferroreview,ferroreview2,hfo2review}. A defining feature of a ferroelectric is the polar distortion --- a manifestation of the collective off-center displacements of atoms~\cite{ferrobook,ferrobook2}. The polar distortion in a ferroelectric governs its properties (e.g., dielectric and piezoelectric responses, ferroelectric hysteresis, and polarization switching dynamics) and functionalities~\cite{ferrobook,ferrobook2,improper1}. Since the 1970s, the concepts of proper, improper and triggered ferroelectrics have been established~\cite{trigger1,trigger3,improper1,ferrobook,ferrobook2}. 
In proper ferroelectrics, polar distortions appear as primary order parameters (OPs)~\footnote{In pseudoproper ferroelectrics, the polar distortion is driven by another distortion, where both distortions have identical symmetry~\cite{ferrobook}. Because of this, we can merge these distortions as an effective polar distortion and treat pseudoproper ferroelectrics as proper ferroelectrics.} and may induce secondary OPs~\cite{ferrobook,ferrobook2,improper1}. Alternatively, polar distortions occur in improper or triggered ferroelectrics as secondary or coupled OPs --- rooted in nonpolar OPs~\cite{trigger1,trigger3,improper1,ferrobook,ferrobook2}. 
These textbook concepts assume that polar distortion has a single proper, improper or triggered feature~\cite{ferrobook,ferrobook2,improper1}, and succeed in elucidating the polar natures in conventional ferroelectrics such as barium titanate. In sharp contrast, applying such concepts to complex ferroelectrics is rather challenging and can yield highly controversial arguments. A notorious example is the polar orthorhombic hafnia compatible with silicon-based electronics~\cite{hfo2cmos1,hfo2cmos2,hfo2cmos3,hfo2cmos4,hfo2cmos5}. As an elusive case, this material was recognized as proper~\cite{hfo2prb}, improper~\cite{hfo2ncomms,hfo2science,hfo2prm} or even triggered~\cite{triggeredhfo2} ferroelectric in literature.

We resolve such an issue by developing a graph theory tailor-made for clarifying the nature of polar distortions in complex ferroelectrics. We emphasize that complex ferroelectrics can exhibit multiple natures among proper, improper and triggered characteristics. We demonstrate the robustness of our theory by working with two representative cases --- that is, perovskite superlattices and polar orthorhombic hafnia --- both of which exhibit elusive polar distortions. Our theory not only correctly captures the mixed proper-improper nature in perovskite superlattices, but also unifies the conflicting arguments in Refs.~\cite{hfo2ncomms,hfo2science,hfo2prm,hfo2prb,triggeredhfo2} regarding polar orthorhombic hafnia (confirming its mixed triggered-improper nature). \\

\noindent
\textit{The selection of the reference phase.} For a ferroelectric, let the symmetry groups of its paraelectric and ferroelectric phases be $G_0$ and $G$, respectively. The ferroelectric phase transition is associated with the symmetry breaking from $G_0$ to $G$. This is described by a subgroup chain as $G_0 \supseteq G_1 \supseteq \cdots \supseteq G$, where $G_1$ and $\cdots$ characterize the symmetry groups of the intermediate phases. The subgroup chains are usually not unique: $G_0 \supseteq G^\prime_1 \supseteq \cdots \supseteq G$ may serve as another subgroup chain. The description of the ferroelectric distortion lies in the selection of a nonpolar reference phase, but this selection has arbitrariness. For instance, $G_0$, $G_1$, $G^\prime_1$, and other states may be selected as reference phases if these states are not polar.

Different choices of the reference phases may lead to inconsistent or contradictory conclusions. To show this, let us imagine that the $G$ phase has a polar $p$ and two nonpolar $(q_1,q_2)$ OPs defined with respect to the $G_0$ phase. We further assume that the ferroelectricity in the $G$ phase is improper and is originating from the $p q_1 q_2$ trilinear coupling. This means that the reversal of the ferroelectric polarization involves the reversal of either $q_1$ or $q_2$ (but not both). The transition from $G_0$ phase to $G$ phase is associated with $G_0  \supseteq G_1^p \supseteq G^{(p,q_1,q_2)}$, $G_0  \supseteq G_1^{q_1} \supseteq G^{(p,q_1,q_2)}$, or $G_0  \supseteq G_1^{q_2} \supseteq G^{(p,q_1,q_2)}$ subgroup chain --- the superscript representing the condensed OPs. By selecting $G_1^{q_1}$ as the reference phase, we conclude that the ferroelectric switching involves the reversal of $p$ and $q_2$ OPs. However, selecting $G_1^{q_2}$ as the reference phase will lead to a very different conclusion, that is, the ferroelectric switching reverses both $p$ and $q_1$. Such contradictory conclusions are due to the local and biased analyses of ferroelectric distortion based on low-symmetric $G_1^{q_1}$ and $G_1^{q_2}$ reference phases. For this reason, we emphasize the necessity to select the highest-symmetric $G_0$ state as the reference phase. This will enable a global description of ferroelectric distortions and avoid inconsistent or contradictory analyses.\\

\begin{figure}[h!]
\centering
\includegraphics[width=1\linewidth]{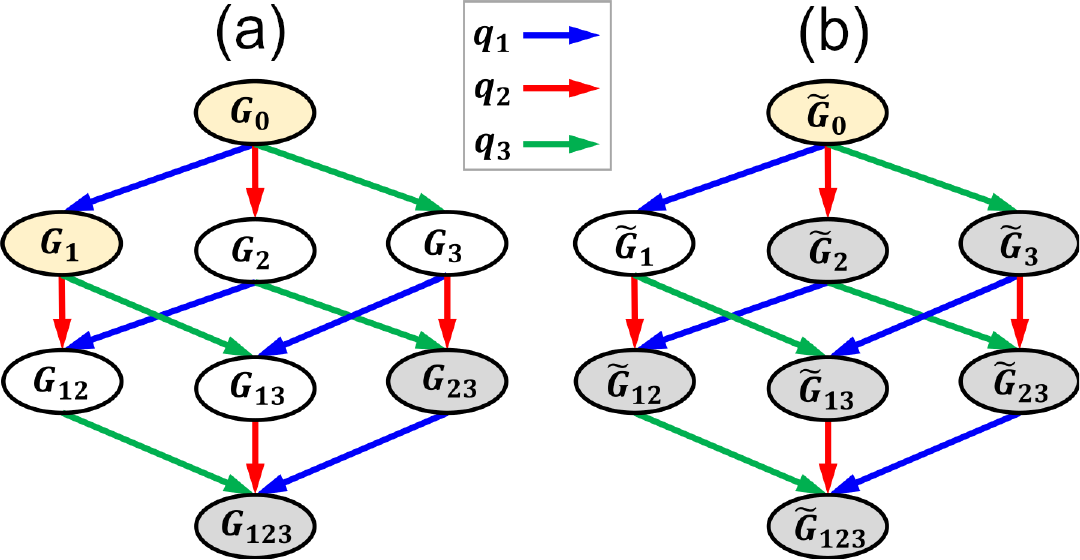}
\caption{\label{fig:graphtheory} Sketches of nonpolar hierarchy graph (a) and polar hierarchy graph (b). The set associated with each vertex (being omitted) is indicated by arrows pointing to that vertex. For instance, the vertexes marked by $G_{13}$ and $G_0$ have sets of $\{ q_1, q_3 \}$ and $\emptyset$, respectively. In panels (a) and (b), the vertexes with yellow (gray) color share identical crystal structures as that for $G_0$ ($G_{123}$); the vertexes with white color have crystal structures that are not equivalent to each other. We further assume that the $\tilde{G}_1$ vertex has a polar space group, vertexes with gray color have polar space groups, and other vertexes have nonpolar space groups.}
\end{figure}

\noindent
\textit{Proper and improper components.} Our discussion is built upon the notions of proper and improper components. To introduce these notions, we start from a ferroelectric with $n$ nonpolar OPs, and such nonpolar OPs form a set 
$Q=\{q_i~|~i=1,2,\cdots,n\}$. 
The ferroelectric also has a polar distortion $p$ and a ferroelectric polarization $P$. We say that this ferroelectric contains a proper component if superimposing $p$ distortion on the reference phase yields a nonzero electric polarization $P_0$ (i.e., $p$ being a primary OP~\footnote{Our work focuses on the ground state structural properties of ferroelectrics. The primary and non-primary OPs are referred to as stable and unstable modes, respectively.}). In the ferroelectric phase, the $P_0$ polarization may be suppressed by some nonpolar OPs and enhanced by the others. Therefore, $P$ usually differs from $P_0$, and the proximity between $|P_0|$ and $|P|$ provides a ``measurement'' of the proper component. Motivated by this, we propose the proper membership coefficient $\eta$ as
\begin{align}\label{eq:fuzzyproper}
    \eta=(|P|\wedge |P_0|)/(|P|\vee |P_0|),
\end{align}
where $\wedge$ ($\vee$) selects the minimum (maximum) value between $|P|$ and $|P_0|$. At the probability level, a larger $\eta$ implies that $P$ is more likely inherited from $P_0$.

When superimposing several nonpolar distortions on the reference phase, the symmetry of such a phase may be broken. If the symmetry breaking is compatible with a polar distortion, the ferroelectric contains improper components. In the following, we develop a graph theory that determines whether a ferroelectric contains improper components. Superimposing one or several nonpolar OPs on the reference phase results in $2^n$ different distortion states that are represented by the elements in $\mathcal{P}(Q)$, the power set of $Q$ given by
\begin{align}\label{eq:powerset}
    \mathcal{P}(Q) & = \{\tilde{Q}~|~\tilde{Q}\subseteq  Q\} =    
    \Big\{ \emptyset, \{q_1\}, \{q_2\}, \cdots, \{q_n\}, \nonumber \\    
  &  \{q_1, q_2\},  \{q_1, q_3\}, \cdots \{q_{n-1}, q_n\}, \{q_1, q_2, q_3 \}, \\
  & \{q_1, q_2, q_4 \}, \cdots, \{q_{n-2}, q_{n-1}, q_n\}, \cdots, Q\Big\},  \nonumber
\end{align}
where $\emptyset$ empty set means that no nonpolar OPs are superimposed on the reference phase, $Q$ means that all the nonpolar OPs are superimposed,  $\{q_i\}$ means that only $q_i$ is superimposed, $\{q_i, q_j\}$ means that both $q_i$ and $q_j$ are superimposed, and so forth. The aforementioned $2^n$ distortion states and their relations can be well described by our proposed nonpolar hierarchy graph. In such a graph, there are $n$ directed edges and $2^n$ vertexes. The edges and vertexes represent nonpolar distortions and nonpolar distortion states, respectively [see Eq.~(\ref{eq:powerset})]. 

For every pair of $(A,B)$ vertexes, we draw a $q_i$ directed edge from $A$ to $B$ if $A \subseteq B$ and $B-A=\{q_i\}$; otherwise, the $A$ and $B$ vertexes are not connected by any edges. We term the $\emptyset$ and $Q$ vertexes as the start and end vertexes, respectively. Each vertex is valued via the following procedures: (i) we create the structural distortions in the reference phase according to the vertex, (ii) we perform first-principles structural relaxations for the initially created crystal structure, and (iii) we assign the vertex to the relaxed crystal structure and the resulting space group~\footnote{We may employ the graph-theoretical approach to understand the magnetically induced ferroelectricity in type-II multiferroics. In such cases, we should include magnetic order parameters and work with magnetic space groups.}. Vertexes that have equivalent crystal structures are colored with the same color, while vertexes that have unique crystal structures are uncolored (i.e., with white color). We can deduce that the ferroelectric contains no improper components if there are no vertexes with polar space group. On the contrary, a $\{q_i, q_j, \cdots \}$ vertex with a polar space group indicates that the combination of $\{q_i, q_j, \cdots \}$ nonpolar OPs yields improper polarization. Walking along the directed edges from the $\emptyset$ vertex to a vertex with polar space group yields a path associated with an improper component (termed as ``improper path''). An improper path is irreducible when it contains only one vertex with polar space group. Otherwise, the path can be further shortened as an irreducible path. Each irreducible improper path indicates a combination of nonpolar OPs --- involving a minimal number of OPs --- that drives an improper component.

The interplay among polar and nonpolar OPs can be further understood by constructing the polar hierarchy graph as follows. First, we represent the vertexes for the polar hierarchy graph by the $\{ \{ p \} \cup\tilde{Q}~|~\tilde{Q}\subseteq  Q \}$ set, where $p$ is the polar distortion and $Q=\{q_i~|~i=1,2,\cdots,n\}$. Then, the polar hierarchy graph is readily obtained by following the rules for constructing nonpolar hierarchy graph (see above). 
The comparison between nonpolar and polar hierarchy graphs allows us to explore the nature of polar distortion in a ferroelectric. If nonpolar hierarchy graph contains enough information, the construction of polar hierarchy graph can be omitted.

As a demonstration, we assume that a ferroelectric has nonpolar $\{q_1, q_2, q_3 \}$ OPs and a polar $p$ OP. 
Figure~\ref{fig:graphtheory} schematizes its nonpolar and polar hierarchy graphs, and these graphs deliver abundant information. First, $G_0$ and $G_1$ vertexes have equivalent crystal structure and therefore $q_1$ is not a primary OP. Second, $q_1$ can be stabilized in the presence of $q_2$, $q_3$, or $p$ (see $G_{12}$, $G_{13}$, and $\tilde{G}_{1}$ vertexes). Third, $q_2$ and $q_3$ are primary OPs and the $(q_2,q_3)$ combination yields an improper polarization. Finally, $p$ is not a primary OP and this ferroelectric contains no proper component [see Fig.~\ref{fig:graphtheory}(b)].\\

\begin{figure}[t!]
\centering
\includegraphics[width=1\linewidth]{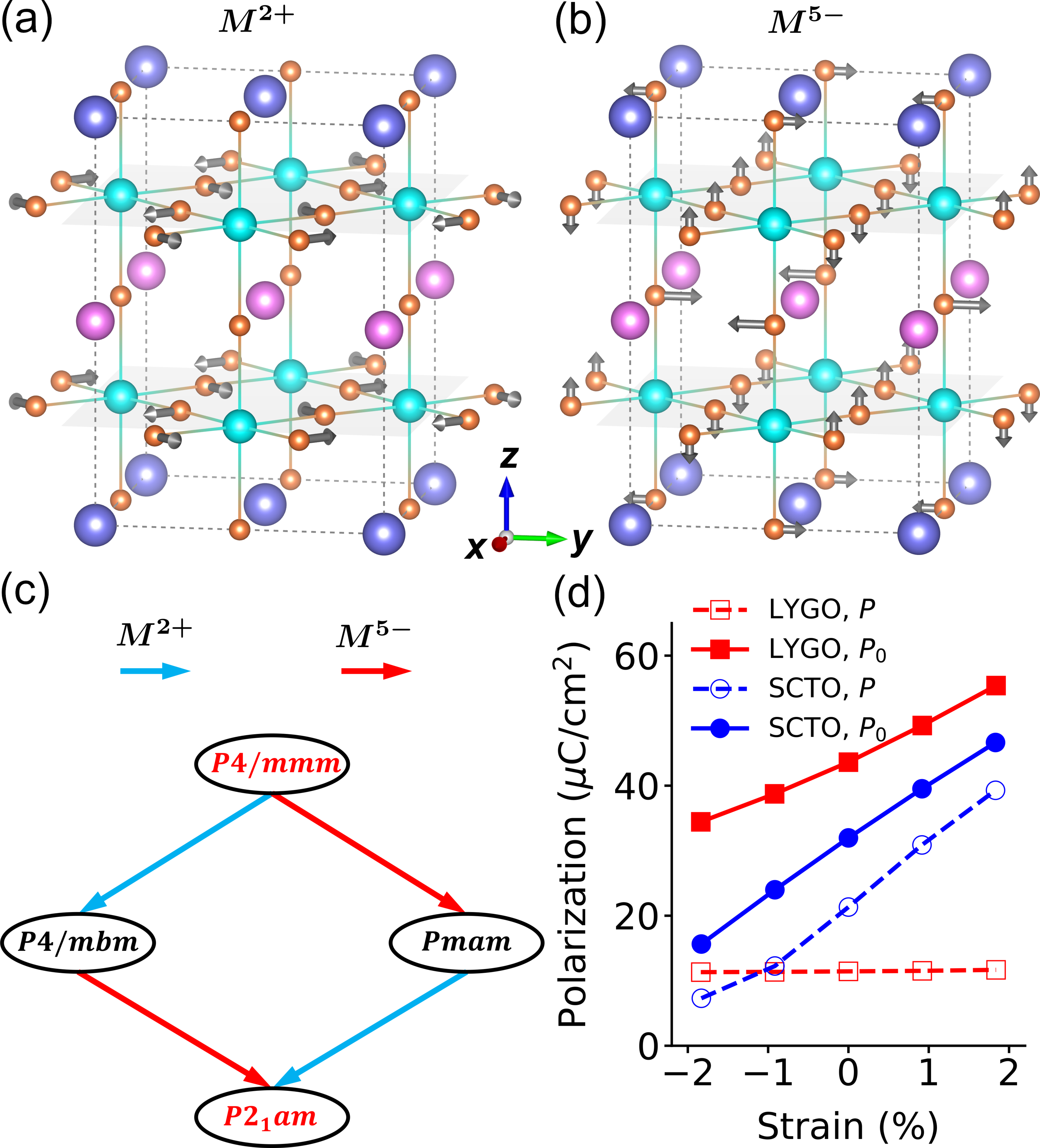}
\caption{\label{fig:superlatt} Ferroelectricity in the $AB$O$_3/A^\prime B$O$_3$ perovskite superlattice. Panels (a) and (b) sketch the nonpolar $M^{2+}$ and $M^{5-}$ distortions, where gray arrows represent ionic motions. The $A$, $A^\prime$, $B$ and O ions are represented by purple, pink, cyan and orange spheres, respectively. In panel (b), the motions of $A$, $A^\prime$ and $B$ ions (being very tiny) are not shown. Panel (c) is the nonpolar hierarchy graph for LaGaO$_3$/YGaO$_3$ and SrTiO$_3$/CaTiO$_3$ superlattices. Panel (d) shows the strain dependent polarizations. The strain for LaGaO$_3$/YGaO$_3$ and SrTiO$_3$/CaTiO$_3$ is defined with respect to $a_0=5.44$~\AA~and $a_0=5.46$~\AA, respectively.}
\end{figure}

\noindent
\textit{Complex ferroelectrics with mixed characteristics.} Following Refs.~\cite{trigger1,trigger3,improper1,ferrobook,ferrobook2,triggeredhfo2,properimproper1}, we revisit the concepts of proper, improper and triggered ferroelectrics. Ideally, conventional ferroelectrics have the following features:

(i) Proper ferroelectric~\cite{ferrobook,ferrobook2,improper1}. A material has a polar distortion as its primary OP and contains no improper components. Nonpolar distortions (if existing) only slightly modify the primary polar OP.

(ii) Improper ferroelectric~\cite{ferrobook,ferrobook2,improper1}. A material has no polar distortion as its primary OP. The polar distortion originates from improper components that are not associated with any secondary nonpolar OPs.

(iii) Triggered ferroelectric~\cite{trigger1}. A material contains neither proper nor improper components. The polar distortion is triggered by one or several nonpolar OPs.

In sharp contrast, complex ferroelectrics usually go beyond these ideal situations and showcase multiple features. Typical examples include:

(iv) Mixed proper-improper ferroelectric~\cite{improper1,ferrobook,ferrobook2,properimproper1}. A material has both proper and improper components.

(v) Mixed proper-triggered ferroelectric~\cite{trigger1,trigger3}. A material has a proper component but has no improper components. The collaborative couplings between polar and nonpolar OPs trigger the simultaneous occurrence of polar distortion and nonpolar OP(s).

(vi) Mixed triggered-improper ferroelectric~\cite{trigger1,improper1,triggeredhfo2}. A material has no proper component. The polar distortion is triggered by improper components, and the nonpolar OPs associated with these improper components are not all primary OPs.

In complex ferroelectrics, boundaries between proper, improper and triggered mechanisms can be very fuzzy. 
Regarding this, our proposed $\eta$ coefficient [see Eq.~(\ref{eq:fuzzyproper})] provides a ``measurement'' of the proper degree of the ferroelectricity. The nature of polar distortion in a ferroelectric with triggered features can be explored by comparing its polar and nonpolar hierarchy graphs. In the following sections, we shall demonstrate the applications of our theory to several complex ferroelectrics. \\

\begin{figure*}[t!]
\centering
\includegraphics[width=1\linewidth]{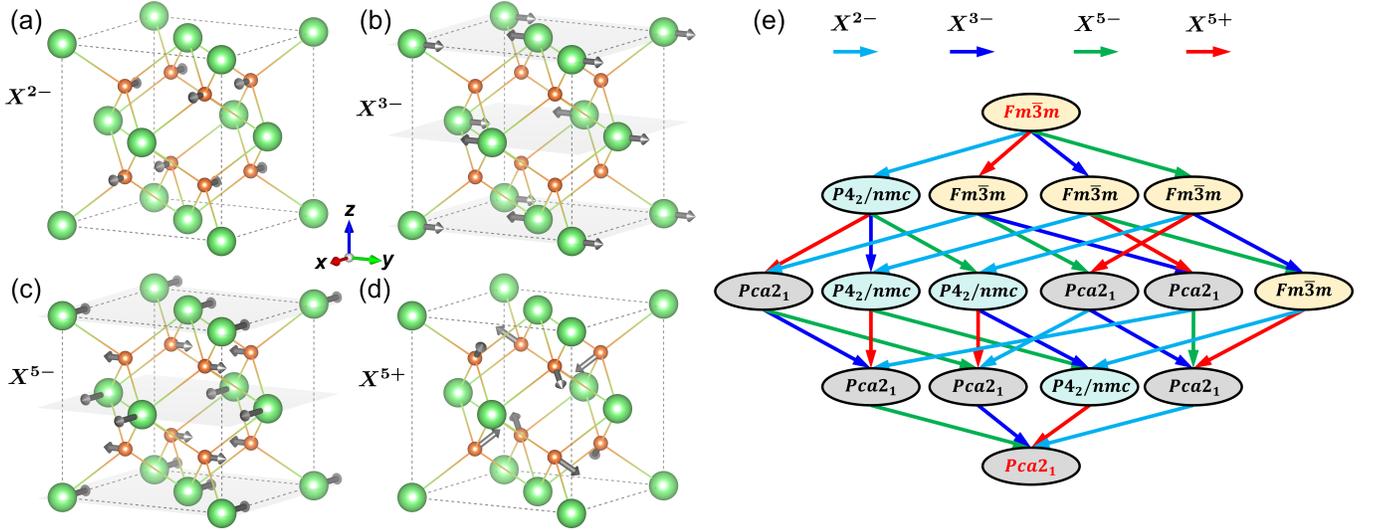}
\caption{\label{fig:hafnia} Ferroelectricity in orthorhombic HfO$_2$. Panels (a)---(d) sketch four nonpolar distortion modes in HfO$_2$, where gray arrows represent ionic motions. The Hf and O ions are represented by green and yellow spheres, respectively. Panel (e) is the phase transition graph for HfO$_2$. Note that the $4_2$ screw axis in $P4_2/nmc$ is along the $x$ direction.}
\end{figure*}

\noindent
\textit{Example I: Ferroelectricity in perovskite superlattices.} Our first example is the short period $AB$O$_3/A^\prime B$O$_3$ perovskite superlattices. Figures~\ref{fig:superlatt}(a)--(b) sketch two primary nonpolar OPs in $AB$O$_3/A^\prime B$O$_3$, namely, $M^{2+}$ and $M^{5-}$ distortions. Such two OPs are defined with respect to the $P4/mmm$ reference phase. The $(M^{2+}, M^{5-})$ combination yields improper polarization, and this is termed as ``hybrid improper ferroelectricity''~\cite{bousquet2008,zhao2014,benedek2011,rondinelli2012,mulder2013,jorge2013}. A typical hybrid improper ferroelectric is the LaGaO$_3$/YGaO$_3$ superlattice (space group being $P2_1am$) that is predicted by Ref.~\cite{rondinelli2012}. Figure~\ref{fig:superlatt}(c) shows the nonpolar hierarchy graph for LaGaO$_3$/YGaO$_3$, where the start and end vertexes represent reference and ferroelectric phases, respectively. We identify two shortest improper paths between the start and end vertexes, and both paths involve the combination of $M^{2+}$ and $M^{5-}$ nonpolar OPs. This confirms that LaGaO$_3$/YGaO$_3$ indeed contains improper component, which is consistent with Ref.~\cite{rondinelli2012}. The comparison between $P2_1am$ and $P4/mmm$ phases yields a $\Gamma^{5-}$ polar distortion. This $\Gamma^{5-}$ distortion is compatible with the $P4/mmm$ reference phase, resulting in an electric polarization of $P_0=51.3~\mu$C/cm$^2$. The ferroelectric $P2_1am$ phase has a polarization of $P=10.9~\mu$C/cm$^2$. According to Eq.~(\ref{eq:fuzzyproper}), the proper membership $\eta$ coefficient is $\sim21\%$ and this implies that the ferroelectricity in LaGaO$_3$/YGaO$_3$ is largely improper.

We move on to examine the ferroelectric nature in the SrTiO$_3$/CaTiO$_3$ superlattice which contains an active ion for proper ferroelectricity (i.e., Ti$^{4+}$). The nonpolar OPs and nonpolar hierarchy graph for SrTiO$_3$/CaTiO$_3$ resemble those for LaGaO$_3$/YGaO$_3$ [see Figs.~\ref{fig:superlatt}(a)--(c)]. Similar to LaGaO$_3$/YGaO$_3$, SrTiO$_3$/CaTiO$_3$ has an improper component. The $P_0$ and $P$ polarization values for SrTiO$_3$/CaTiO$_3$ are $39.6$ and $22.2~\mu$C/cm$^2$, respectively. The $\eta$ membership coefficient for SrTiO$_3$/CaTiO$_3$ reaches $\sim56\%$, and such a value implies that the
ferroelectricity in SrTiO$_3$/CaTiO$_3$ is proper by a slight majority. Figure~\ref{fig:superlatt}(d) shows the $P$ and $P_0$ polarizations for LaGaO$_3$/YGaO$_3$ and SrTiO$_3$/CaTiO$_3$ films that depends on the biaxial strain (in-plane lattice parameter)~\footnote{To mimic a film on a cubic substrate, we set its in-plane lattice parameters equal to each other. In its bulk counterpart, such two lattice parameters may be different. In our cases, LaGaO$_3$/YGaO$_3$ or SrTiO$_3$/CaTiO$_3$ film with zero strain do not exactly correspond to LaGaO$_3$/YGaO$_3$ or SrTiO$_3$/CaTiO$_3$ bulk material.}. In LaGaO$_3$/YGaO$_3$, $P_0$ is obviously enhanced by increasing the in-plane lattice parameter, while $P$ remains nearly unchanged. This implies that the $P_0$ polarization is significantly suppressed by $M^{2+}$ and/or $M^{5-}$ OPs, and the polarization in $P2_1 am$ LaGaO$_3$/YGaO$_3$ is predominantly contributed by the $(M^{2+}, M^{5-})$ combination. As for SrTiO$_3$/CaTiO$_3$, both $P_0$ and $P$ are significantly enhanced by increasing the in-plane lattice parameter. The similar trends for $P_0$ and $P$ suggests that $P$ in $P2_1 am$  SrTiO$_3$/CaTiO$_3$ is mostly inherited from $P_0$. In Fig.~\ref{fig:peroveta} of the End Matter~\footnote{The End Matter section includes our methods and two supplementary figures (citing Refs.~\cite{kresse1996efficient,kresse1999ultrasoft,pbesol,blochl1994projector,berryphase,vtst1,vtst2,pseudo,amplimode1,amplimode2,findsymurl,stokes2005findsym,momma2011vesta,hunter2007matplotlib}).}, we show the $\eta$ coefficients for these two materials as functions of strains. We find that strain only slightly affects
the proper component in LaGaO$_3$/YGaO$_3$ but significantly increases such a component in SrTiO$_3$/CaTiO$_3$.\\

\begin{figure*}[t!]
\centering
\includegraphics[width=1\linewidth]{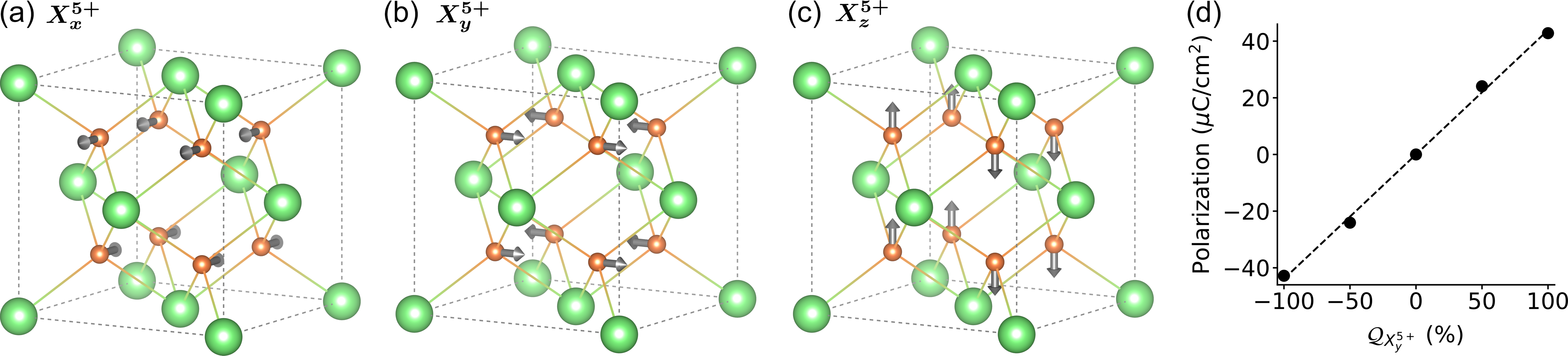}
\caption{\label{fig:ferrohafnia} The ferroelectricity in orthorhombic HfO$_2$ driven by the $X^{5+}$ distortion. Panels (a), (b), and (c) sketch the $X^{5+}_x$, $X^{5+}_y$, and $X^{5+}_z$ modes, respectively. The $X^{5+}$ mode, the compass and the axis labels can be found in Fig.~\ref{fig:hafnia}. Panel (d) shows the polarization of HfO$_2$ as a function of the $X^{5+}_y$ distortion (i.e., $\mathcal{Q}_{X^{5+}_y}$). See \textit{Methods} for the computational details.}
\end{figure*}

\noindent
\textit{Example II: Ferroelectricity in polar orthorhombic hafnia.} The next example is the orthorhombic HfO$_2$ with a  $Pca2_1$ space group. References~\cite{hfo2ncomms,hfo2science,hfo2prm,hfo2prb,triggeredhfo2} select $Fm\bar{3}m$, $P4_2/nmc$ or $Pbcn$ phase as its reference phase. We shall show that selecting the $Fm\bar{3}m$ reference phase (as is done in Refs.~\cite{hfo2ncomms,hfo2science,triggeredhfo2}) can not only interpret the results in Refs.~\cite{hfo2prm,hfo2prb} but also provide more deep understandings of the ferroelectric nature in $Pca2_1$ HfO$_2$. With respect to $Fm\bar{3}m$ phase, we identify four nonpolar OPs in $Pca2_1$ HfO$_2$ and demonstrate them in Figs.~\ref{fig:hafnia}(a)--(d). Compared with the $Fm\bar{3}m$ reference phase, $Pca2_1$ HfO$_2$ contains a $\Gamma^{4-}$ polar distortion. Such a distortion is not compatible with the $Fm\bar{3}m$ phase (i.e., $P_0=0$). The nonpolar hierarchy graph of HfO$_2$ is shown in Fig.~\ref{fig:hafnia}(e).
The end vertex is associated with the polar $Pca2_1$ phase which implies that HfO$_2$ contains improper components. From the graph, we identify the irreducible improper paths that involve $(X^{5+},X^{2-})$, $(X^{5+},X^{3-})$, and $(X^{5+},X^{5-})$ OPs. Among the four nonpolar distortions, only the $X^{2-}$ mode is the primary OP, while the other three are not. The role of the $X^{2-}$ OP is very special in $Pca2_1$ HfO$_2$. Superimposing this OP on the $Fm\bar{3}m$ phase yields the $P4_2/nmc$ phase. Our phonon calculation confirms that the $P4_2/nmc$ phase has no imaginary phonon modes, which suggests the triggered ferroelectric nature (coinciding with Ref.~\cite{triggeredhfo2}). We further find that removing $X^{5+}$ edges breaks the connectivity between the start and end vertexes. Hence, $Pca2_1$ HfO$_2$ is a mixed triggered-improper ferroelectric, and the $(X^{2-},X^{5+})$ combination is critical for creating the ferroelectric polarization.

The nonpolar $X^{5+}$ mode in Fig.~\ref{fig:hafnia}(d) involves noncollinear motions of O ions. Such motions can be further decomposed into $X^{5+}_x$, $X^{5+}_y$ and $X^{5+}_z$ collinear modes [see Figs.~\ref{fig:ferrohafnia}(a)--(c)]. To compare with Refs.~\cite{hfo2prm,hfo2prb}, we construct a more complicated nonpolar hierarchy graph by working with $X^{2-}$, $X^{5-}$, $X^{5+}_x$, $X^{5+}_y$ and $X^{5+}_z$ nonpolar OPs (see Fig.~\ref{fig:hafniafull} of the End Matter). Using $P4_2/nmc$ phase as a reference phase, Ref.~\cite{hfo2prm} argues that the polarization in $Pca2_1$ HfO$_2$ arises from the combination of $Y_{4-}$ and $Y_{2+}$ OPs which, in essence, correspond to our $X^{5+}_x$ and $X^{5+}_z$ OPs, respectively. As shown in Fig.~\ref{fig:hafniafull}, the $(X^{2-}, X^{5+}_x, X^{5+}_z)$ combination yields the polar $Pca2_1$ phase. Recalling that the $P4_2/nmc$ phase is created by condensing the $X^{2-}$ mode, the $(X^{5+}_x, X^{5+}_z)$ combination yields an improper component in the $P4_2/nmc$ phase --- coinciding with Ref.~\cite{hfo2prm}. Different from Ref.~\cite{hfo2prm}, Ref.~\cite{hfo2prb} selects the $Pcan$ state as a reference phase, and conclude that the polarization in $Pca2_1$ HfO$_2$ is proper. Our analysis suggests that condensations of $X^{2-}$, $X^{5-}$ and $X^{5+}_z$ OPs result in the $Pcan$ state. Taking the $Pcan$ phase as a reference, the polarization in $Pca2_1$ HfO$_2$ is associated with $X^{5+}_y$ or $X^{5+}_x$ nonpolar OP. As a illustration, we show that $X^{5+}_y$ can induce polarization in the $Pcan$ state [see Fig.~\ref{fig:ferrohafnia}(d)]. To summarize this paragraph, our theory enables a global description of the ferroelectricity in $Pca2_1$ HfO$_2$, and this unifies the arguments in Refs.~\cite{hfo2ncomms,hfo2science,hfo2prm,hfo2prb,triggeredhfo2}.\\

\noindent
\textit{Perspective.} We explicitly establish the concepts of complex ferroelectrics with multiple natures among proper, improper and triggered characteristics. The necessity for introducing such concepts is justified by that complex ferroelectrics can yield exotic phenomena that are unlikely to occur in conventional ferroelectrics. Very recently, Ref.~\cite{properimproper1} predicts the coexisted proper and improper characteristics and the resultant ferrielectric-like unusual hysteresis in CsNbW$_2$O$_9$. Another example is the unusual hysteresis feature arisen in $Pca2_1$ HfO$_2$ with mixed triggered-improper characteristics~\cite{triggeredhfo2}.

The development of Landau theory is the textbook approach for understanding the polar distortion in ferroelectrics~\cite{ferrobook,ferrobook2}. However, due to incomplete and biased analyses of energetic couplings, such an approach may lead to conflicting pictures for complex ferroelectrics. In this regard, our graph theory provides a straightforward way for clarifying the polar distortion natures in complex ferroelectrics. This will definitely deepen our knowledge on ferroelectric physics and benefit the discovery of novel ferroelectrics and related phenomena.\\

\noindent
\textit{Acknowledgements.} We thank the support from the National Natural Science Foundation of China with Grants No.~12274174, No.~52288102, No.~52090024, and No.~12034009. L.B. acknowledges the Vannevar Bush Faculty Fellowship (VBFF) grant No. N00014-20-1-2834 from the Department of Defense and award No. DMR-1906383 from the National Science Foundation AMASE-i Program (MonArk NSF Quantum Foundry).\\

%

\vspace{\baselineskip}
\vspace{\baselineskip}

\onecolumngrid

\appendix

\section{End Matter}

\noindent
\textit{Methods.} We use the VASP~\cite{kresse1996efficient,kresse1999ultrasoft} code and employ the PBEsol~\cite{pbesol} functional (PAW~\cite{blochl1994projector} potentials) to perform our first-principles numerical simulations. We set the kinetic energy cutoff as 600 eV, and solve the following electronic configurations: $(5s,5p,5d,6s)$ for La, $(4s,4p,4d,5s)$ for Y, $(3d,4s,4p)$ for Ga, $(4s,4p,5s)$ for Sr, $(3s,3p,4s)$ for Ca, $(3s,3p,3d,4s)$ for Ti, $(5s,5p,5d,6s)$ for Hf, and $(2s,2p)$ for O. We use a $8\times8\times6$ $k$-point mesh for LaGaO$_3$/YGaO$_3$ and SrTiO$_3$/CaTiO$_3$ and a $10\times10\times10$ $k$-point mesh for HfO$_2$; such meshes correspond to the lattices demonstrated in Figs.~\ref{fig:superlatt}(a) and~\ref{fig:hafnia}(a). We perform the structural relaxations by setting the force convergence criterion as 0.001 eV/\AA. When mimicking a strained film on a cubic substrate, we fix the in-plane lattice vectors of the film to $(a,0,0)$ and $(0,a,0)$ --- $a$ being adjusted according to the substrate lattice constant. Then, we relax the out-of-plane lattice vector together with the atomic positions. We compute the electric polarization values for LaGaO$_3$/YGaO$_3$, SrTiO$_3$/CaTiO$_3$, and HfO$_2$ by the Berry phase method~\cite{berryphase} with the help of the VTST tookit~\cite{vtst1,vtst2}. We carry out our symmetry analysis by the PSEUDO~\cite{pseudo}, AMPLIMODES~\cite{amplimode1,amplimode2} and FINDSYM~\cite{findsymurl,stokes2005findsym} tookits, visualize our crystal structures and atomic motions by the VESTA~\cite{momma2011vesta} code, and prepare our plots by the Matplotlib library~\cite{hunter2007matplotlib}.

When calculating Fig.~\ref{fig:ferrohafnia}(d), the amplitude of the $X^{5+}_x$ mode is fixed to 0, amplitudes of $X^{2-}$ and $X^{5-}$ modes are fixed to their values in bulk $Pca2_1$ HfO$_2$, and the amplitude of the $X^{5+}_y$ mode is fixed to $-100\%$, $-50\%$, $0\%$, $50\%$, or $100\%$ of that in bulk $Pca2_1$ HfO$_2$; furthermore, an initial $X^{5+}_z$ distortion is given according to its amplitude in bulk $Pca2_1$ HfO$_2$, and $\Gamma^{4-}$, $X^{3-}$ and  $X^{5+}_z$ modes are free to relax. Such calculations are achieved by the ``selective dynamics'' strategy (see \url{https://www.vasp.at/wiki/index.php/POSCAR}) by fixing the fractional coordinates of several ions and relaxing the other degrees of freedom including the lattice vectors. Therefore, the fixed amplitudes are performed at the level of fixed ionic fractional coordinates.\\

\begin{figure*}[h!]
\centering
\includegraphics[width=0.9\linewidth]{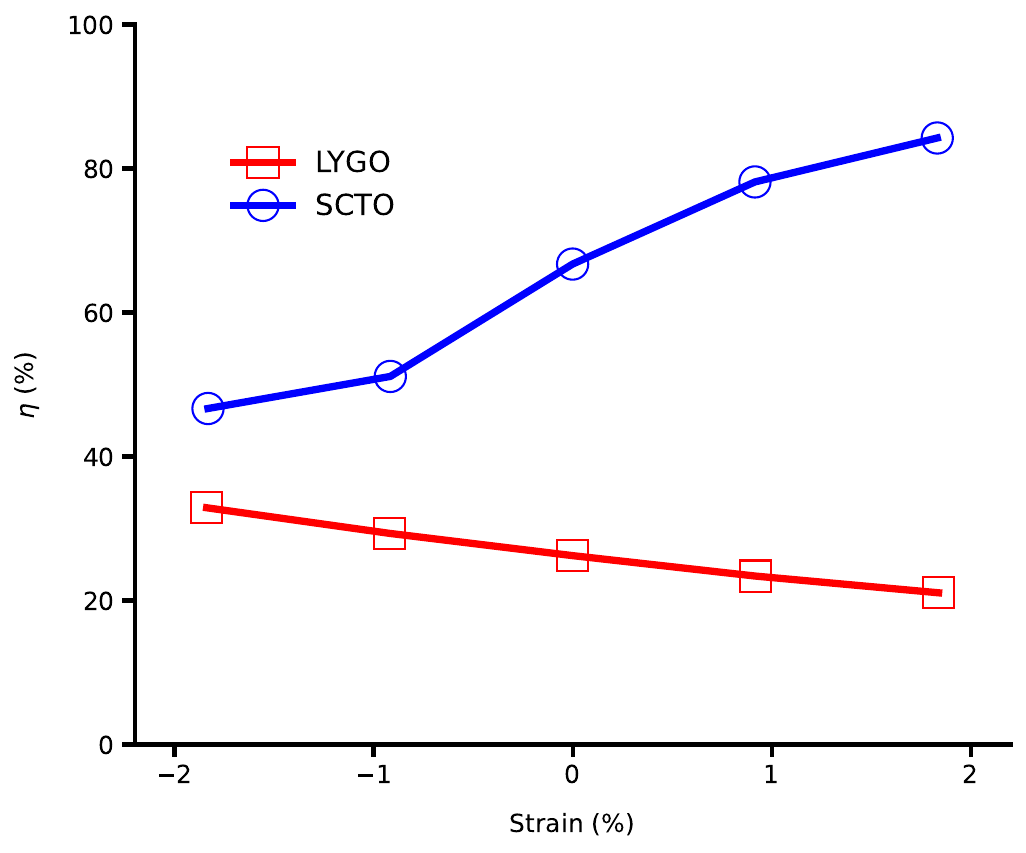}
\caption{\label{fig:peroveta} The strain-dependent membership coefficients for LaGaO$_3$/YGaO$_3$ and SrTiO$_3$/CaTiO$_3$ superlattices.}
\end{figure*}

\begin{figure*}[t!]
\centering
\includegraphics[width=1\linewidth]{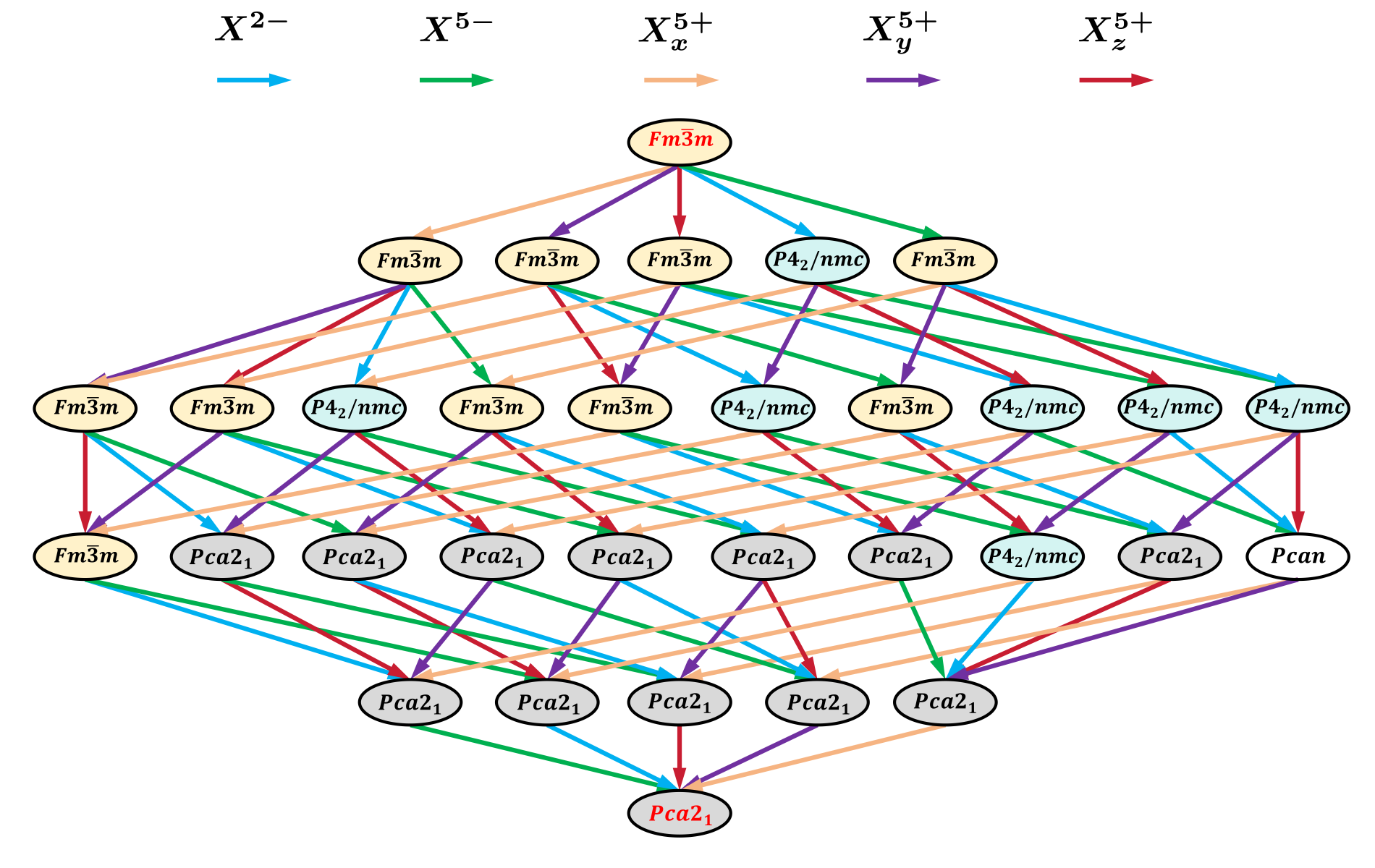}
\caption{\label{fig:hafniafull} The nonpolar hierarchy graph for HfO$_2$ constructed with respect to $X^{2-}$, $X^{5-}$, $X^{5+}_x$, $X^{5+}_y$, and $X^{5+}_z$ distortion modes. Note that the $4_2$ screw axis in $P4_2/nmc$ is along the $x$ direction. Furthermore, the $Pcan$ is the non-standard setting for the $Pbcn$ space group.}
\end{figure*}

\noindent
\textit{Two supplementary figures.} We also provide two supplementary figures to support our discussion in the Main Text. Figures~\ref{fig:peroveta} shows the $\eta$ membership coefficients for strained LaGaO$_3$/YGaO$_3$ and SrTiO$_3$/CaTiO$_3$, while Figure~\ref{fig:hafniafull} demonstrates a nonpolar hierarchy graph for $Pca2_1$ HfO$_2$ constructed with respect to five nonpolar OPs.

\end{document}